# Measuring Cognitive Load of Software Developers Based on Nasal Skin Temperature

Keitaro NAKASAI[†], Shin KOMEDA[††], Masateru TSUNODA[††], and Masayuki KASHIMA[†††]

**SUMMARY** It has recently become increasingly important to measure the cognitive load of developers. This is because continuing with a development under a high cognitive load may cause human errors. Therefore, to automatically measure the cognitive load, existing studies have used biometric measures such as brain waves and the heart rate. However, developers are often required to equip certain devices when measuring them, and can therefore be physically burdened. In this study, we evaluated the feasibility of non-invasive biometric measures based on the nasal skin temperature. The nasal skin temperature has been widely used in other fields to measure mental status. In the present experiment, the subjects created small Java programs, and we estimated their cognitive load using the proposed biometric measures based on the nasal skin temperature. As a result, the proposed biometric measures were shown to be more effective than non-biometric measures. Hence, biometric measures based on nasal skin temperature are promising for estimating the cognitive load of developers.

*Key words: Nasal temperature, mental workload, NASA-TLX, affective engineering*

## 1. Introduction

Using biometric measures, some studies have recently attempted to measure the cognitive load of developers during software development [4]. Biometric measures are derived by measuring the physiological states of the developers, such as their brain waves. Cognitive load refers to the extent of mental burden of a person engaged in a task. When the cognitive load of a task is high, it suggests that the person is coping with mentally difficult tasks. For instance, when work requires significant calculations within a short period, the cognitive load becomes high.

A continual high cognitive load may cause human errors [8]. If we can identify such a high cognitive load of developers, we can decrease the risk of errors by warning developers of such a load. In addition, we can measure the skill of a developer based on the cognitive load [5]. For instance, when developers solve the same programming problem, the cognitive load of some developers is lower than that of the others. In this case, the programming skills of the former are regarded as relatively higher.

It is not easy for other people to recognize the cognitive load of a developer based on appearance. If biometric measures can be used to measure the cognitive load of a developer, they can be measured automatically and objectively. As a result, we can utilize the cognitive load to achieve the goals described above. Biometric measures such as brain waves, heart rate, and fMRI have been used to measure the cognitive load of software developers [4].

However, to conduct such measurements, developers are often required to equip certain devices, and therefore may experience a certain amount of physical burden. For instance, to measure brain waves, a conductive gel is placed on the electrodes, and the subjects should wash their hair after a measurement.

The goal of our study was to evaluate the feasibility of new biometric measures that can estimate a cognitive load without placing a physical burden on the developers. We propose the use of noninvasive biometric measures based on nasal skin temperature. The nasal skin temperature of an individual can be measured using thermography. When thermography is used, developers are not required to install any device. The nasal skin temperature has been widely used in other fields to measure mental status [1][3][6][10][15].

## 2. Related Work

Gonçales et al. [4] surveyed the cognitive load of different software developers. As sensors for measuring cognitive load, electroencephalograms (EEGs), eye-trackers, functional magnetic resonance imaging (fMRI), functional near-infrared spectroscopy (fNIRS), electromyography (EMG), and multimodal sensors have been used. EEGs are used to measure brain waves. Based on blood flow in the brain, fMRI shows which parts of the brain are activated. Similarly, fNIRS uses near-infrared light to clarify the active parts of the brain. In addition, EMG is used to clarify the activities of the muscles. Developers should equip certain devices to conduct such measurements, which are therefore regarded as invasive biometric measures.

Some eye-tracker devices are noninvasive [11]. Note that our purpose is to show the feasibility of a new noninvasive approach, but not to replace the nasal skin temperature with the eye gaze. Both are non-invasive approaches, and we can therefore combine them to measure a cognitive load. An evaluation of the effect of such combination is a topic for future research.

Multimodal sensors combine different sensors to measure a cognitive load. For instance, skin temperature and heart rate variability are used in combination, and some devices are required to measure such metrics. For example, to measure the skin temperature, Schaule at al. [13] used a bracelet-like device, and similar studies have not used the nasal skin temperature or noninvasive biometric measures.

---

[†] The author is with NIT, Kagoshima College, Kirishima-shi, 899-5193 Japan.
[††] The authors are with Kindai University, Higashiosaka-shi, 577-8502 Japan.
[†††] The author is with Kagoshima University, Kagoshima-shi, 890-8580 Japan.

## 3. Nasal Skin Temperature

When the sympathetic nervous system is active (i.e., the mental status of a person is active), blood flow and skin temperature decrease. There are peripheral blood vessels just under the nose, and the nasal skin temperature is significantly affected by blood flow. Hence, using the nasal skin temperature, we can indirectly observe the activity of the sympathetic nervous system [1][10]. Specifically, when the nasal skin temperature is low, the mental status of a person is active.

The nasal skin temperature has been widely used in other fields to measure a mental status [1][3][6][10][15]. The advantage of nasal skin temperature is its non-invasive biometric measurement [1][10]. Previous studies have used this approach to measure the cognitive load. To continuously increase their mental workload, Mizuno et al. [6] proposed assigning a 3-min mental arithmetic task to their subjects six times. As the results indicate, the nasal skin temperature can appropriately measure the cognitive load. Likewise, Study [1] measured the cognitive load using the nasal skin temperature while reading texts of various complexities.

The nasal skin temperature has also been used to measure the arousal level. For example, Nozawa et al. [10] used nasal skin temperature to measure the arousal level. Similarly, Diaz-Piedra et al. [3] used the nasal skin temperature to measure the arousal level during a 2-h simulated driving experiment. Moreover, nasal skin temperature is used to estimate the stress in children [15].

 Based on past studies in other fields, the nasal skin temperature is expected to be useful in measuring the cognitive load during a software development. Unlike the mental arithmetic applied in [6], programming consists of the following phases:

- Planning the structure of the source code.
- Inputting the source code based on the plan.
- Removing defects included in the code.

The extent of the cognitive load in each phase might be different; hence, it is unclear how nasal skin temperature is effective in measuring the cognitive load during programming.

## 4. Metrics for Estimation of Cognitive Load

An estimation model of cognitive load is typically built based on the following procedure:

1. A subject engages in certain tasks, during which the biometric measurements of the subject are collected.
2. After the task, the cognitive load is measured using a self-assessment questionnaire [1].
3. A prediction model is then built, treating the cognitive load measured in step 2 and the biometric measures collected in step 1 as dependent and independent variables, respectively.

The task load index (NASA-TLX) is often used to measure cognitive load [1][17]. NASA-TLX is a questionnaire that measures the mental demand, physical demand, temporal demand, own performance, effort, and frustration. To measure the cognitive load of software developers, 83% of studies have posed programming tasks for developers [4]. Specifically, the use of change tasks, comprehension tasks, and a combination of both tasks have been posed.

Because the forehead skin temperature is not significantly affected by blood flow [10], and the difference between the forehead and nasal skin temperatures signifies the status of the sympathetic nervous system, we calculated such difference to allow using the nasal skin temperature as a biometric measure, as shown in Figure 1. This difference, which we refer to in the following as the NST, has frequently been applied in previous studies, including [1][3][10]. Using the NST, we defined the following biometric measures:

- **WNST**: NST during a task minus NST while at rest
- **WMAX**: Maximum value of WNST
- **WAVE**: Average value of WNST
- **WSUM**: Sum of WNST

In our experiment, the NST differed among the subjects. That is, the NST of some of the subjects was high, whereas that of the others was low. To normalize the differences, we defined the WNST. As described in Section 3, the programming consists of different phases, and the cognitive load may differ among phases. Hence, we assumed that the WNST varies during a task, and WMAX is defined to focus on the maximum load on the task. WAVE is defined to consider the average cognitive load during a task. WSUM considers the total cognitive load during the task.

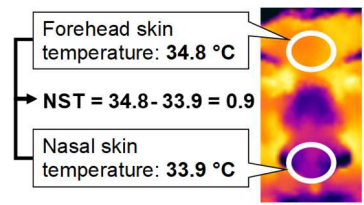

**Fig. 1** Example metrics defined in this study

| Task time (min.) | 2 | 4 | 6 | 8 |
|---|---|---|---|---|
| Forehead temperature | 32.2 | 32.1 | 32.3 | 32.1 |
| Nasal temperature | 34.9 | 34.7 | 34.8 | 34.7 |
| NST during a task | 2.7 | 2.6 | 2.5 | 2.6 |
| WNST | 1.2 | 1.1 | 1.0 | 1.1 |

WMAX  WSUM = 4.4  WAVE = 1.1
NST during rest = 1.5

**Fig. 2** Example metrics defined in this study

**Table 1** Adjusted $R^2$ of each model

| Candidates of Independent variables | Dependent variables | | | |
|---|---|---|---|---|
| | Mental demand | Own performance | Effort | Frustration level |
| Time | 0.20 | -0.06 | 0.30 | 0.01 |
| Time, WMAX, WAVE, and WSUM | 0.27 | **0.64** | **0.56** | 0.43 |

**Table 2** Standardized partial regression coefficients of each model

| Independent variable | Dependent variables | | | |
|---|---|---|---|---|
| | Mental demand | Own performance | Effort | Frustration level |
| WMAX | 0.46 | - | - | 0.68 |
| WAVG | - | -1.10 | 0.31 | -1.32 |
| WSUM | - | 1.22 | 0.63 | - |
| Time | -0.80 | -0.67 | -1.25 | - |

Figure 2 shows an example of these metrics. For example, the forehead and nasal skin temperatures were recorded every 2 min, and the metrics were calculated at each point in time. In the example, we assumed that the NST at rest was 1.5.

## 5. Experiment

In the experiment, the subjects engaged in the following three programming tasks:

- Preliminary: (printing "hello world") to learn the experiment environment.
- Easy: (printing a multiplication table using a for loop).
- Difficult: (counting the digits of the given numbers).

We selected the tasks from Aizu Online Judge [2]. The order of easy and difficult tasks was randomly changed for each subject. The subjects rested for 3 min before each task to drop the NST. The rest time was the same as that in a previous study [6].

The participants edited the programs on a web browser. The accuracy of the program was tested using JUnit. When they clicked the execute button, the execution result of the program and the output of JUnit were shown in the browser. We recorded each task time (i.e., from the time required to show the specifications of the program to the time needed to pass the Junit test).

After each task, the subjects answered a questionnaire that simplified NASA-TLX [14] to measure the cognitive load (see Section 4). NASA-TLX measures not only the mental load but also the physical load, as described in Section 4. Hence, past studies such as that described in [17] used some of the metrics. Similarly, we used mental demand, own performance, effort, and frustration level.

Seven undergraduate (fourth year) and master course students majoring in computer science were used as the subjects. We used a FLIR T530 for thermography. The distance between the thermograph and the subjects was approximately 1 m, and the emissivity was set to 0.98. Because we assume that the programming phases described in Section 3 are not frequently switched to other phases, and the NST gradually changes, we calculated the metrics described in Section 4 every 2 min.

Similar to previous studies [7], to evaluate the proposed metrics, we compared them with a non-biometric measure. As a non-biometric measure, we used the task time to solve the problems. Similar to our study, previous studies such as those in [7] and [9] mainly focused on evaluating the feasibility of biometric measures and did not compare them with other biometric measures.

## 5. Results

The average task time was 6.9 min for an easy task, and was 13.1 min for a difficult task. The average WNST was 0.70, and the maximum and minimum values were 1.3 and 0.26 respectively.

We created two types of linear regression model to evaluate the explanatory power of the proposed metrics. We set WMAX, WAVE, WSUM, and the task time as the explanatory variables of the multivariate linear regression models. In the existing approach, we set only the task time as an explanatory variable. We applied a log transformation to the task time. Mental demand, own performance, effort, and frustration level measured by the NASA-TLX were set as the dependent variables of each model.

There were 14 data points because each subject engaged in two tasks. We applied a stepwise variable selection method to remove superfluous variables. To consider multicollinearity, when the tolerance was smaller than 1.0, the variable was not added to the model during selection.

Table 1 shows the adjusted $R^2$ values for each model. The explanatory power of the model increased according to the number of independent variables. The adjusted $R^2$ was used to evaluate the explanatory power, avoiding the influence of the number. When the value was greater than 0.5, the model explained the dependent variables well.

All adjusted $R^2$ values of the models based on the proposed metrics were higher than those of the non-biometric measure (benchmark). In addition, the $R^2$ values were larger than 0.5 when the independent variables were own performance and effort. Therefore, the proposed metrics are more effective than the benchmark, and can be used to estimate the own performance and effort.

Table 2 shows the standardized partial regression coefficients of each model. In the table, "-" in the cells indicate that the variable was not included in the model. When the values of each dependent variable are small, the cognitive load is regarded to be high. The partial coefficient of time was negative, indicating that when time is long, the values of the independent variables are low (i.e., the cognitive load is high), which is a reasonable result.

The partial coefficients of WMAX and WSUM are positive. This means that when the NST is high, the values of the independent variables are high and the cognitive load is low. This is also reasonable because when the NST is high, the mental status of the subjects is not extremely active. WSUM is considered the most effective because the adjusted $R^2$ of models that include WSUM is larger than 0.5.

In addition, the correlation coefficients are not small. However, the signs of the partial coefficients of the WAVG are inconsistent among the models. Further experiments are needed to clarify the role of WAVG in estimating the cognitive load.

## 6. Threats to Validity

The participants in the experiment were not professional software developers. Previous studies have shown that students instead of professionals can be used in such experiments [12]. Therefore, we believe that the results of our study would not have been extremely different had professionals participated in our experiment. The recruitment of professional developers will be an area of future study, however.

The number of subjects was also small. However, it is not easy to increase the number of subjects in this type of study. For example, 10 subjects were included in the studies described in [6], [7], and [9]. However, we acknowledge that the number of subjects should be increased to enhance the reliability of the results.

In addition, in our experiment, we used only a small number of specifications owing to time limitations [16]. However, it is difficult to use realistic programs in such experiments. For instance, one study [9] used codes whose lines ranged from 17 to 32. However, the size must be considered when interpreting the results.

## 7. Conclusion

We evaluated the feasibility of non-invasive biometric measures, WMAX, WAVG, and WSUM, which are based on the nasal skin temperature. Such temperature has been widely used in other fields for mental status measurements. In the feasibility evaluation, we conducted an experiment in which the subjects engaged in three programming tasks. After the task, the cognitive load was measured using the NASA-TLX questionnaire.

We estimated the cognitive load based on the proposed metrics using a multivariate linear regression. In addition, as a benchmark, we estimated the performance based on a non-biometric measure (i.e., task time). As a result, the explanatory power of the proposed metrics was larger than that of the non-biometric measure, and own performance and effort as measured by NASA-TLX were sufficiently explained through the proposed model.

One of our future studies will be to increase the number of subjects to enhance the reliability of the results. In addition, we will evaluate the effect of the combination of our metrics with other noninvasive biometric measures (e.g., metrics based on eye gaze) for estimating the cognitive load of the developers.

**Acknowledgments**

This research was partially supported by JSPS (Grants-in-Aid for Scientific Research (C) (No. 21K11840)).